\documentclass[12pt]{article} 
\usepackage[top=30mm, bottom=40mm, left=25mm, right=25mm]{geometry}

\usepackage{graphicx}
\usepackage{textcomp} 

\usepackage{epstopdf}

\usepackage{bm}

\usepackage{latexsym}

\usepackage[mathcal]{euscript}
\usepackage{slashed}

\usepackage{indentfirst}

\usepackage{color}
\usepackage{amssymb}
\usepackage{amsmath}
\newcommand{\dt}{\dot}

\usepackage{hyperref}

\numberwithin{equation}{section}

\newcommand{\be}{\begin{equation}}
\newcommand{\ee}{\end{equation}}
\newcommand{\bea}{\begin{eqnarray}}
\newcommand{\eea}{\end{eqnarray}}

\newcommand{\al}{\alpha}
\renewcommand{\d}{\delta}
\newcommand{\e}{\epsilon}
\newcommand{\G}{\Gamma}
\newcommand{\g}{\gamma}
\renewcommand{\k}{\kappa}
\newcommand{\La}{\Lambda}
\newcommand{\la}{\lambda}
\newcommand{\m}{\mu}

\newcommand{\Om}{\Omega}

\newcommand{\s}{\sigma}
\renewcommand{\t}{\theta}

\newcommand{\hlf}{\frac{1}{2}}

\newcommand{\non}{\nonumber}

\newcommand{\p}{\partial}

\newcommand{\R}{\mathbb{R}}

\newcommand{\w}{\wedge}
\newcommand{\Z}{\mathbb{Z}}

\newcommand{\lp}{\left(}
\newcommand{\rp}{\right)}
\newcommand{\ls}{\left[}
\newcommand{\rs}{\right]}

\newcommand{\hph}[1]{{\hphantom{#1}}}
\newcommand{\ov}[1]{{\overline{#1}}}

\newcommand{\newbullet}{\bullet}

\title{
{\bf Abelian Tensor Hierarchy in 4D, $\bm {N=1}$ Superspace}
}
\author{
Katrin Becker, 
Melanie Becker, 
William D. Linch \textsc{iii}, 
and
Daniel Robbins
}
\date{
} 

\begin{document}

\maketitle

\vspace*{-65mm}
\begin{flushright}
{MI-TH-1605}
\end{flushright}
\vspace*{+40mm}

\begin{center}
{\em
George P. and Cynthia W. Mitchell Institute for
Fundamental Physics and Astronomy, \\
Texas A\&{}M University,
College Station, TX 77843 USA.
}\\
\end{center}

\vspace*{15mm}

\begin{abstract}
With the goal of constructing the supersymmetric action for all fields, massless and massive, obtained by Kaluza-Klein compactification from type II theory or M-theory in a closed form, we
embed the (Abelian) tensor hierarchy of $p$-forms in four-dimensional, $N=1$ superspace and construct its Chern-Simons-like invariants.
When specialized to the case in which the tensors arise from a higher-dimensional theory, the invariants may be interpreted as higher-dimensional Chern-Simons forms reduced to four dimensions.
As an application of the formalism, we construct the eleven-dimensional Chern-Simons form in terms of four-dimensional, $N=1$ superfields.
\end{abstract}

\thispagestyle{empty}

\newpage
\tableofcontents
\thispagestyle{empty}
\newpage
\setcounter{page}1

\section{Introduction}

Kaluza-Klein theory was discovered long ago \cite{Kaluza:1921tu, Klein:1926tv} in an attempt to unify the only known forces at that time, electromagnetism and gravity.
By postulating a fifth dimension of space-time, the electromagnetic field is considered as a component of gravity, rather than a fundamental force.
Since then, a multitude of new ideas have been added to the original five-dimensional proposal, among them supersymmetry and the extension to
all possible space-time dimensions and compactification manifolds. Some beautiful and almost successful attempts to describe our four-dimensional world have appeared in the literature \cite{Witten:1981me}.

No matter which scenario is being considered though, they all have one common feature: the appearance of additional massless scalars (not present in electromagnetism nor gravity) and an infinite tower of massive Kaluza-Klein states. Initially, physicists tried to deal with this by truncating the higher dimensional theory in order to find models resembling our four-dimensional world, but often such truncations were not consistent (see e.g.~\cite{Duff:1984hn}).
In the modern approach to Kaluza-Klein theory (pioneered in refs. \cite{Scherk:1975fm, Cremmer:1975sj, Cremmer:1976ir})
extra dimensions and the corresponding massive harmonics are treated as physical and not merely as mathematical structures. In the meantime a precise definition of a consistent truncation
has been found (see e.g.~\cite{Duff:1986hr}).
Some of these truncations involve a finite number of massive states \cite{Maldacena:2008wh, Gauntlett:2009zw}, which become relevant e.g.\ in the context of non-relativistic conformal field theories.

More recently, in the context of type IIA and M-theory compactifications to two, three, and four dimensions on $G_2$ and $Spin(7)$ structure manifolds, the conditions for having a supersymmetric vacuum were derived from the dynamics of massive Kaluza-Klein modes \cite{Becker:2014rea}. In particular, certain interactions in space-time were inferred and used to determine the $F$- and $D$-term conditions for unbroken supersymmetry.
Classically these conditions mean that the
$G_2$ or $Spin(7)$ structure manifolds have a $G_2$ or $Spin(7)$ holonomy metric. Explicitly, a superpotential was conjectured and the invariance of the space-time action under
gauge transformations of the M-theory three-form required the associated moment map to vanish.
In the $G_2$ case
these two conditions imply the existence of a closed three-form and closed four-form.
Moreover, classically the K\"ahler potential for chiral multiplets is related to the volume of the internal space which implies that the three-form is the Hodge dual of the four-form. Consequently the internal space has a $G_2$ holonomy metric. Beyond the classical limit there still exists a closed three-form and a closed four-form but they are no longer Hodge dual to each other.

In ref.~\cite{Becker:2014uya} we started constructing explicitly the space-time theory obtained when reducing (super-) gravity and certain matter fields (including $p$-form tensor fields) to any number of space-time dimensions. The type of theories considered is quite general and includes type II string theory and M-theory reduced to two, three, and four dimensions.
The actions obtained in ref.~\cite{Becker:2014rea} for M-theory compactified to four dimensions involved bosonic fields only,
and the aim of our program is to describe the manifestly supersymmetric completion.
To achieve this,
the fields and interactions described in ref.~\cite{Becker:2014rea} will be assembled into superfields of $d=4$ and $N=1$ supersymmetry.

The approach we are using is quite general and actually not new. An early
publication writing a higher dimensional theory in lower dimensional superspace is ref.~\cite{Marcus:1983wb} in which the formulation of ten-dimensional supersymmetric Yang-Mills theory in four-dimensional, $N=1$ superspace was presented.\footnote{This result was rediscovered more recently in ref.~\cite{ArkaniHamed:2001tb}.}
The inclusion of gravity has (to our knowledge) not been worked out and remains a challenging problem. Even before coupling to gravity it would be interesting to work out the three-dimensional version of the tensor hierarchy presented in this paper in
superspace. This would be a step in the direction of writing
the three-dimensional quantum field theory obtained by compactifying type IIA/IIB theories to three dimensions in three-dimensional superspace. 

The actions of the type considered in ref.~\cite{Becker:2014rea} result from splitting the spacetime coordinates into two parts and are, being a rewriting of the original theory, more general than a compactification. Nevertheless, many compactification phenomena will have analogues in such a splitting, an important one of which is the existence of a ``gravitational tensor hierarchy'' \cite{deWit:2005hv}. This consists of a collection of $p$-form gauge fields coming from the dimensionally reduced component forms of the original supergravity theory organized into a hierarchy and coupled to non-Abelian gauge fields resulting from the vector-like part of the dimensionally reduced graviton. Any complete, manifestly 4D, $N=1$ description of eleven-dimensional supergravity will have a superspace analogue of such a non-abelian tensor hierarchy.

Apart from their appearance in maximal supergravities, tensor hierarchies may be considered in their own right as an extension of charged matter fields to forms of degree higher than 1. In six dimensions, this idea has been used in attempts to construct conformal theories with $N=(1,0)$ supersymmetry \cite{Samtleben:2011fj,Samtleben:2012mi,Samtleben:2012fb}. In such models, the forms do not (necessarily) arise from the reduction of differential forms in higher dimensions and it is, therefore, useful to construct such tensor models in a formalism that does not commit to a differential-geometric origin.

This paper represents a modest step the direction of constructing the actions with local supersymmetry in superspace.
We present a model
consisting of anti-symmetric tensor fields subjected to some symmetries to which we will refer as the ``Abelian
tensor hierarchy''. We present the bosonic form and the corresponding superspace version (with global four-dimensional, $N=1$ supersymmetry).
In a forthcoming publication this is generalized to a non-Abelian tensor hierarchy by gauging \cite{BBLR2}. The construction of the locally supersymmetric generalization is in progress \cite{BBLR3}.

\section{Bosonic Tensor Hierarchy}

In this section we present the bosonic Abelian tensor hierarchy. It consists of a series of $p$-form fields in $d$-dimensional space-time taking values in some vector spaces $V_p$. The dimension of $V_p$ is the number of $p$-forms, which could be infinite. We take the space-time metric to be flat and subject the $p$-form tensor fields to a set of Abelian
gauge transformations. These
gauge transformations are inspired by, but not identical to, those obtained when compactifying the eleven-dimensional three-form to four space-time dimensions. We show how the system obtained from dimensional reduction arises as a special case of the more general Abelian tensor hierarchy.

\subsection{Potentials and Gauge Transformations}

Consider a collection of real scalars, one-forms, two-forms, and so on up to $p$-forms in $d$ dimensions. In this section we keep $d$ arbitrary, while in the rest of this paper we take $d=4$.  We write
\be
\phi^A,\qquad \phi_a^I,\qquad\phi_{ab}^M,\qquad\phi_{abc}^S,\qquad\phi_{abcd}^X,\qquad \dots
\ee
where $A$ runs over the set of scalars, $I$ runs over the vectors, $M$ over the two-forms, and so on.
In the remainder of this section we also use an alternative indexing for the fields in some equations, writing $I_0,I_1,\cdots,I_p,\cdots$ instead of $A,I,\cdots$.  This allows us to write more general formulae.  In equations without explicit space-time indices we use a subscript $[p]$ to make clear that the given object is a $p$-form, i.e.\ $\phi^{I_p}_{[p]}$.
The fields  $\phi^{I_p}_{a_1 \dots a_p}$ are functions taking values in a real vector
space $V_p$ with $I_p=1,\dots, \mathrm{dim}(V_p)$. In the concrete examples discussed in this paper $V_p$ will be
the space of differential forms of some degree, $\Omega^{n-p}(M)$, on a manifold $M$. But for now we keep matters general and do
not specialize to this case.

For each $p>0$ there is a gauge transformation parameterized by a differential $(p-1)$-form $\La^{I_p}_{[p-1]}$, which generates Abelian $p$-form transformations. In addition, there is a shift by the parameter $\La^{I_{p+1}}_{[p]}$.  For instance
\begin{equation}
\begin{split}
& \d\phi^A=\lp q^{(0)}\rp^A_I\La^I,\\
& \d\phi^I_a=\p_a\La^I+\lp q^{(1)}\rp^I_M\La^M_a,\\
& \d\phi^M_{ab}=2\p_{[a}\La^M_{b]}+\lp q^{(2)}\rp^M_S\La^S_{ab},
\end{split}
\end{equation}
or in general
\be
\d\phi^{I_p}_{a_1\cdots a_p}=p\p_{[a_1}\La^{I_p}_{a_2\cdots a_p]}+\lp q^{(p)}\rp^{I_p}_{J_{p+1}}\La^{J_{p+1}}_{a_1\cdots a_p},
\ee
where $(q^{(p)})^{I_p}_{J_{p+1}}$ are linear maps
\begin{equation}
q^{(p)}: V_{p+1} \to V_{p}.
\end{equation}
In differential form notation,
\be\label{e4}
\d\phi^{I_p}_{[p]}=d\La^{I_p}_{[p-1]}+\left( q^{(p)} \cdot \La_{[p]} \right)^{I_p}.
\ee
Here, $d$ denote the exterior derivative and we introduced the notation
\begin{equation}
\left( q^{(p)} \cdot \La_{[p]} \right)^{I_p}=\left(  q^{(p)}\right)^{I_p}_{J_{p+1}} \La^{J_{p+1}}_{[p]} .
\end{equation}

We define the field strengths
\begin{equation}
\label{eq:BosonicFieldStrengths}
F^{I_p}_{[p+1]}=d\phi^{I_p}_{[p]}-\lp q^{(p)}\cdot  \phi_{[p+1]}\rp^{I_p},
\end{equation}
which satisfy
\be
\d F^{I_p}_{[p+1]}=-\lp q^{(p)}\cdot q^{(p+1)}\cdot \La_{[p+1]}\rp^{I_p}.
\ee
In order for the field strengths to be gauge invariant, we thus require that
\be\label{e9}
\lp q^{(p)}\cdot q^{(p+1)}\rp^{I_p} _{K_{p+2}}=0, \qquad \forall p \geq 0.
\ee
It is then natural to interpret $q$ as the boundary operator for a chain complex $V_\newbullet$,
\be
\label{eq:VComplex}
V_\newbullet=\left\{ \cdots\stackrel{q^{(p+1)}}{\longrightarrow}V_{p+1}\stackrel{q^{(p)}}
{\longrightarrow}V_p\stackrel{q^{(p-1)}}{\longrightarrow}V_{p-1}
\stackrel{q^{(p-2)}}{\longrightarrow}\cdots\stackrel{q^{(0)}}{\longrightarrow}V_0\right\}.
\ee
Because of eqn.\ (\ref{e9}),
\begin{equation}
{\rm im} q^{(p+1)}\subseteq {\rm ker} q^{(p)},
\end{equation}
but in general there is no equality. It is this mismatch which gives rise to interesting physical quantities, as we explain in detail in section \ref{subsec:Homology}.

In addition to $V_\newbullet$, we have the $d$-dimensional de Rham complex,
\be
\Om^\newbullet\left(\R^{d-1,1}\right)=\left\{
\Om^{0}\stackrel{d}{\longrightarrow}
\Om^{1} \stackrel{d}{\longrightarrow}\cdots \stackrel{d}{\longrightarrow}
\Om^{p}
\stackrel{d}{\longrightarrow}\cdots\right\}.
\ee
Then the gauge fields $\phi_{[p]}$ take values in $\Om^{p}\otimes V_p$, the gauge parameters $\La_{[p-1]}$ in $\Om^{p-1}\otimes V_p$, and the field strengths $F_{[p+1]}$ in $\Om^{p+1}\otimes V_p$.
The field strengths satisfy the Bianchi identities,
\be
\label{eq:BosonicBianchis}
dF^{I_p}_{[p+1]}=-\lp q^{(p)}\cdot F_{[p+2]}\rp^{I_p} .
\ee

There is one more phenomenon that we will need which is the extension of the complex (\ref{eq:VComplex}) one step further to the right, i.e. a new space $V_{-1}$ and a linear operator $q^{(-1)}:V_0\longrightarrow V_{-1}$ satisfying $q^{(-1)}\cdot q^{(0)}=0$
\be
\label{eq:ExtendedVComplex}
V_\newbullet = \left\{ \cdots\stackrel{q^{(0)}}{\longrightarrow}V_0\stackrel{q^{(-1)}}{\longrightarrow}V_{-1}\right\}.
\ee
In terms of matrices, if we let $Z$ index $V_{-1}$, then we require
\be
\lp q^{(-1)}\rp^Z_A\lp q^{(0)}\rp^A_I=0.
\ee
With this understood, we can naturally define a new ``field strength'',
\be
\label{eq:F0Def}
F^Z_{[0]}=-\lp q^{(0)}\rp^Z_A\phi^A_{[0]}.
\ee
This is a gauge-invariant linear combination of the scalars $\phi_{[0]}^A$ which is handed to us in the case that the complex is extended as in (\ref{eq:ExtendedVComplex}).  Note that since there are no $(-1)$-forms on $\R^4$ , i.e.\ $\Omega_{-1}(\R^4)=0$, there is no corresponding gauge field $\phi_{[-1]}^Z$, and thus (\ref{eq:F0Def}) is completely consistent with (\ref{eq:BosonicFieldStrengths}).  Also, $F_{[0]}^Z$ satisfies a Bianchi just like (\ref{eq:BosonicBianchis})
\be
dF^Z_{[0]}=-\lp q^{(0)}\rp^Z_AF^A_{[1]}.
\ee

\subsection{Example from Dimensional Reduction}

As an example of how this construction can arise naturally,
consider a $D$-dimensional theory that has an $n$-form potential field $C_{[n]}$.  A good example to keep in mind is eleven-dimensional supergravity, with $D=11$ and $n=3$, or its close cousin with $D=5$ and $n=1$. Let $d$ be an integer $d<D$.
We can formally split the $D$ coordinates into $d$ space-time coordinates $x^a$ and $D-d$ coordinates $y^i$ which are treated as internal labels. The resulting theory is formulated in $d$ space-time dimensions.

For simplicity, we take the space-time to be a product $\R^{d-1,1 }\times M$, where $M$ is a $(D-d)$-dimensional manifold.
The $n$-form $C_{[n]}$ then decomposes into pieces
\begin{equation}\label{e15}
C_{a_1\cdots a_p i_1\cdots i_k}, \qquad p\leq d,\quad k\leq D-d,\quad p+k=n.
\end{equation}

Explicitly, we have
\be
V_p \cong \Om^{n-p}(M),
\ee
the space of differential $(n-p)$-forms on $M$. The boundary case $V_{-1}$ needed to accommodate $F_{[0]}$ is then $\Om^{n+1}(M)$. In general, $\Om^{n-p}(M)$ are infinite-dimensional vector spaces.
Consequently, an infinite number of $d$-dimensional fields can arise.
Some fields are massless and arise from harmonic forms on $M$. If $M$
is compact, the number of such fields is finite. However, there is also an infinite
set of massive fields.

The decomposition of $C_{[n]}$ in eqn. (\ref{e15}) reflects the K\"unneth decomposition
\be
\label{eq:KunnethDecomposition}
\Om^n(\R^{d-1,1}\times M)\cong\bigoplus_p \Om^p(\R^{d-1,1})\otimes\Om^{n-p}(M) .
\ee
The operators $q^{(p)}$ are also easy to identify. They are the exterior derivative $d_M$ of $M$, acting on $\Om^{n-p}(M)$.  The field strength $F_{[p+1]}$
is the projection of $dC_{[n]}$ onto the appropriate summand in eqn. (\ref{eq:KunnethDecomposition}).

It can be instructive to formulate these matters a bit more explicitly.
Differential $p$-forms in space-time, $\phi^{I_p}_{[p]}$ are labeled by a multi-index
\begin{equation}
I_p = (i_1, \dots, i_{n-p};y),
\end{equation}
which includes $(n-p)$ indices on $M$, as well as the dependence on the ``internal'' coordinate $y$.
Thus, for this example
\be
\phi_{[p]}^{I_p}=C_{a_1\cdots a_pi_1\cdots i_{n-p}}(x,y).
\ee
(Note that although we wrote the $\phi^I$s previously with an upper field index $I$, in this context it is more natural to use lowered indices.)  The contraction of these field indices then includes an integral over the position $y$. This is called deWitt notation. For example, given two fields
\begin{equation}
 u^{(i;y)}(x)=u^{i}(x,y)
 ~~~\mathrm{and}~~~
 v_{(i ;y)}(x)=v_{i}(x,y),
\end{equation}
then
\be
u^{(i;y)}(x)v_{(i;y)}(x')=\sum_i \int_Md^{D-d}y\,u^{i}(x,y)v_{i}(x',y),
\ee
and analogously for fields carrying any number of indices along $M$.

We take $q$ to be the set of operators
\be
\label{eq:DimRedq}
\lp q^{(p)}\rp_{(i_1\cdots i_{n-p};y)}^{\hph{(i_1\cdots i_{n-p};y)}(j_1\cdots j_{n-p-1};y')}=\lp -1\rp^{n-1}\lp n-p\rp\d^{[j_1}_{[i_1}\cdots\d^{j_{n-p-1}]}_{i_{n-p-1}}\p_{i_{n-p}]}\d(y-y').
\ee
It is not difficult to verify that $q^2=0$.  Indeed, $q$ applies $d_M$ so, being a bit schematic,
\be\label{e23}
q^{(p)}\cdot\phi_{[p+1]}=d_M\phi_{[p+1]}.
\ee
Here $\phi_{[p+1]}$ is a differential $(n-p-1)$-form in $M$ and $d_M$ increases the internal degree by one, leaving the space-time degree fixed.  Then both sides of eqn. (\ref{e23}) have space-time degree $p+1$ and internal degree $n-p$.

The gauge transformations and field strengths in eqns. (\ref{e4}) and (\ref{eq:BosonicFieldStrengths}) become
\begin{equation}
\begin{split}
& \d C_{a_1\cdots a_pi_1\cdots i_{n-p}}=p\p_{[a_1}\La_{a_2\cdots a_p]i_1\cdots i_{n-p}}+\lp -1\rp^p\lp n-p\rp\p_{[i_1}\La_{|a_1\cdots a_p|i_2\cdots i_{n-p}]},\\
& F_{a_1\cdots a_{p+1}i_1\cdots i_{n-p}}=\lp p+1\rp\p_{[a_1}C_{a_2\cdots a_{p+1}]i_1\cdots i_{n-p}}+\lp -1\rp^{p+1}\lp n-p\rp\p_{[i_1}C_{|a_1\cdots a_{p+1}|i_2\cdots i_{n-p-1}]}.
\end{split}
\end{equation}
These correspond to the decomposition of the eleven-dimensional equations $\d C = d \La$ and
$F = d C$ in accordance with eqn. (\ref{eq:KunnethDecomposition}).

\subsection{Massless Spectrum and Chain Homology}
\label{subsec:Homology}

Given a chain complex like (\ref{eq:VComplex}) or (\ref{eq:ExtendedVComplex}), it is natural to consider the associated homology groups $H_p(V_\newbullet)=\operatorname{ker}(q^{(p-1)})/\operatorname{im}(q^{(p)})$.  What is the physical significance of this construction?  Any field that lies in the image of $q$ is pure gauge and can be fixed to zero,
while another field that is not in the kernel of $q$ gets a mass via the St\"uckelberg mechanism.  It is sometimes said that the latter field has ``eaten" the former and become massive.
The homology of the chain complex measures what is left, i.e.\ the fields that are in the kernel of $q$ but not in the image of $q$, and these are precisely the fields that remain massless.  Let's see how this works in more detail.

To start with, we will build a basis for each $V_p$.  We could denote an initial basis as $\{e_{I_p}\}$, so that we have expansions like
\be
\phi_{[p]}=\sum_{I_p}\phi^{I_p}_{[p]}e_{I_p}.
\ee
Now we would like to decompose our space further using the boundary maps $q^{(p)}$, and change basis appropriately.  We start at the top of the complex, with $p=d$.  For $V_d$ we first construct a basis $\{a_{\al_d}\}$ for the subspace $\operatorname{ker}(q^{(d-1)})\subseteq V_d$.  Then we complete this with vectors $\{b_{\m_d}\}$ to get a basis for all of $V_d$.  Of course, this new basis could be expanded in terms of the old one $\{e_{I_d}\}$,
\be
a_{\al_d}=\sum_{I_d}a^{I_d}_{\al_d}e_{I_d},\qquad b_{\m_d}=\sum_{I_d}b^{I_d}_{\m_d}e_{I_d}.
\ee
Next, for each $p<d$ we build a basis with three disjoint collections of vectors.  First we take the collection $\{c_{\m_{p+1}}=(q^{(p)}\cdot b_{\m_{p+1}})\}$, where very explicitly,
\be
c_{\m_{p+1}}^{I_p}=\lp q^{(p)}\rp^{I_p}_{J_{p+1}}b^{J_{p+1}}_{\m_{p+1}},\qquad c_{\m_{p+1}}=\sum_{I_p}c^{I_p}_{\m_{p+1}}e_{I_p}.
\ee
These are a basis for $\operatorname{im}(q^{(p)})$.  Next, since $\operatorname{im}(q^{(p)})$ is a subspace of $\operatorname{ker}(q^{(p-1)})$, we can complete this with vectors $\{a_{\al_p}\}$ to get a basis for all of $\operatorname{ker}(q^{(p-1)})$.  Finally, we complete this to a basis for all of $V_p$ with a collection of vectors $\{b_{\m_p}\}$.  Now any vector in $V_p$ can be expanded, for instance
\be
\phi_{[p]}=\sum_{\m_{p+1}}\phi^{\m_{p+1}}_{[p]}c_{\m_{p+1}}+\sum_{\al_p}\phi^{\al_p}_{[p]}a_{\al_p}+\sum_{\m_p}\phi^{\m_p}_{[p]}b_{\m_p}.
\ee
Denote the subspaces of $V_p$ spanned by the $\{c_{\m_{p+1}}\}$, $\{a_{\al_p}\}$, and $\{b_{\m_p}\}$ by $C_p$, $A_p$, and $B_p$, respectively.  Then we have
\begin{align}
\begin{array}{rcccccc}
V_p&\cong & \operatorname{im}(q^{(p)}) & \oplus & \operatorname{ker}(q^{(p-1)})/\operatorname{im}(q^{(p)}) & \oplus & V_p/\operatorname{ker}(q^{(p-1)})\\[0.2cm]
 &\cong& C_p & \oplus & A_p & \oplus & B_p.
\end{array}
\end{align}
In particular, we have $\operatorname{im}(q^{(p)})\cong C_p$, $\operatorname{ker}(q^{(p-1)})\cong C_p\oplus A_p$, and
\be
q^{(p)}\Big|_{B_{p+1}}:B_{p+1}\stackrel{\sim}{\longrightarrow}C_p
\ee
is an isomorphism, and the homology is given by
\be
H_p(V_\newbullet)\cong A_p.
\ee

We now plug these into some of our formulae.  The variations become
\begin{align}
\begin{array}{lrcl}
C_p:& \qquad\d\phi^{\m_{p+1}}_{[p]} &=& d\La^{\m_{p+1}}_{[p-1]}+\La^{\m_{p+1}}_{[p]},\\[0.2cm]
A_p:& \qquad\d\phi^{\al_p}_{[p]} &=& d\La^{\al_p}_{[p-1]},\\[0.2cm]
B_p:& \qquad\d\phi^{\m_p}_{[p]} &=& d\La^{\m_p}_{[p-1]}.
\end{array}
\end{align}
We can use the shift symmetry in the first line to set $\phi^{\m_{p+1}}_{[p]}=0$, thus fixing the gauge symmetry parameterized by $\La^{\m_{p+1}}_{[p]}$.  There is still, in principle, a symmetry corresponding to $\La^{\m_{p+1}}_{[p-1]}$, but it must be compensated by $\La^{\m_{p+1}}_{[p]}=-d\La^{\m_{p+1}}_{[p-1]}$ in order to preserve our gauge choice and nothing transforms under this combination.  After implementing this gauge fixing for each $p$, we are left with the second and third groups of potentials, taking values in $A_p$ and $B_p$, respectively.  The $\phi^{\al_p}_{[p]}$ still enjoy their gauge transformations, parameterized by $\La^{\al_p}_{[p-1]}$, but they are standard Abelian transformations with no extra shift.  The $\phi^{\m_p}_{[p]}$ no longer transform, since we fixed their gauge transformations.

After gauge fixing, the field strengths thus break into
\bea
F^{\m_{p+1}}_{[p+1]} &=& -\phi^{\m_{p+1}}_{[p+1]},\\
F^{\al_p}_{[p+1]} &=& d\phi^{\al_p}_{[p]},\\
F^{\m_p}_{[p+1]} &=& d\phi^{\m_p}_{[p]}.
\eea
Recall that for $p=-1$ there is no potential, and in this case the only non-vanishing components of the field strength are of the first type (taking values in $C_{-1}\cong B_0$),
\be
F^{\m_0}_{[0]}= -\phi^{\m_0}_{[0]}.
\ee

We see immediately that the potentials valued in $A_p\cong H_p(V_\newbullet)$ appear only differentiated ($d\phi^{\al_p}_{[p]}$) and hence these fields must remain massless.  On the other hand the remaining fields $\phi^{\m_p}_{[p]}$ that take values in $B_p$ do appear undifferentiated inside of $F^{\m_p}_{[p]}$.  To make it explicit that these fields are truly massive, and to compute the details of their spectrum, requires some further assumption about the precise form of the kinetic terms.  However, there is nothing protecting them from being massive, and indeed if the kinetic terms have a reasonably standard form
\be
\mathcal{L}_{\mathrm{kin}}=\sum_{p=-1}^3G^{(p)}_{I_pJ_p}F^{I_p}_{[p+1]}\w\ast F^{J_p}_{[p+1]},
\ee
(where $\ast$ is the space-time Hodge duality operator, so $\ast F^{J_p}_{[p+1]}$ is a $(3-p)$-form in space-time and $G^{(p)}$ is some non-degenerate metric on $V_p$), then mass terms arise explicitly from the pieces where we restrict $G^{(p)}$ to $C_p\otimes C_p$.

In the dimensional reduction case, this story translates to something more familiar.  In particular, as mentioned before, the chain complex $V_\newbullet$ is just the co-chain complex $\Om^{n-\newbullet}(M)$, with $q$ being identified with the de Rham exterior derivative $d_M$ on $M$.  The homology groups of $V_\newbullet$ are just the real de Rham cohomology groups of $M$:
\be
H_p(V_\newbullet) \cong H^{n-p}(M,\R).
\ee
When translated into this context, the discussion above amounts to the statements 
\begin{enumerate}
\item We can gauge away the fields corresponding to exact forms on the internal space.
\item The massless fields correspond to the above cohomology groups (with harmonic forms typically used as representatives for the cohomology classes).
\item The fields corresponding to non-closed internal forms generally get masses.  In a spectral decomposition, the masses (squared) would be given in terms of the eigenvalues of the Laplacian operator acting on $\Om^\newbullet(M)$.
\end{enumerate}
We now turn to the superfield embedding of this hierarchy of bosonic $p$-forms.

\section{Superfields}
\label{sec:Superfields}

In this section we will specialize to $d=4$ and embed the hierarchy of bosonic $p$-forms into superfields.
For clarity, we give more conventional names to our potentials:  Instead of $\phi^A_{[0]}$, we will have an axion $a^A$.  $\phi^I_{[1]\,a}$, $\phi^M_{[2]\,ab}$, $\phi^S_{[3]\,abc}$ and $\phi^X_{[4]\,abcd}$ will become $A^I_a$, $B^M_{ab}$, $C^S_{abc}$ and $D^X_{abcd}$ respectively.  The gauge parameters are denoted by $\La_{[p-1]}$, and the field strengths are denoted by $F_{[p+1]}$, including the case $p=-1$.
Our superspace conventions are those of ref.~\cite{Wess:1992cp}, which mostly agree with those of ref.~\cite{Buchbinder:1998qv}; some useful conventions are summarized in Appendix \ref{app:Conventions}.

\begin{table}[h!]
\centering
\label{tab:table1}
\begin{tabular}{|c |c |c| c| c|}
\hline
0-forms & 1-forms & 2-forms & 3-forms & 4-forms\\
\hline
&&&&\\
$a^A$ & $F_a^A$ & $\partial_{[a} F^A_{b]}=0$&                             & \\
&&&&\\
      & $A_a^I$ & $F_{ab}^I$                & $\partial_{[a} F_{bc]}^I=0$  & \\
      &&&&\\
      &         & $B_{ab}^M$                & $F_{abc}^M$                  &$\partial_{[a} F_{bcd]}^M=0 $ \\
      &&&&\\
& & & $C_{abc}^S$ & $ F_{abcd}^S $ \\
&&&&\\
& & & & $D_{abcd}^X$ \\
&&&&\\
\hline
\end{tabular}
\caption{Bosonic fields of the four dimensional Abelian tensor hierarchy.
The potentials are on the main diagonal, field strengths in the next and the Bianchi identities in
the upper diagonal.
Space-time
$j$-forms are in the $j$-th column. When embedded into superfields entries in the same column appear in the same type of superfield. Table 2 displays the superspace version of this table.  }
\end{table}

\begin{table}
\centering
\label{tab:table1}
\begin{tabular}{|c |c |c| c| c|}
\hline
0-forms & 1-forms & 2-forms & 3-forms & 4-forms\\
\hline
&&&&\\
$\Phi^A$ & $F^A$ & $\bar D^2 D_\al F^A=0$&                             & \\
&&&&\\
      & $V^I$ & $W_\al^I$                & $D^\al  W_\al^I - \bar D_{\dot \al}W^{\dot \al}  =0$  & \\
      &&&&\\
         &         & $\Sigma_\al^M$                & $H^M$                  &${\bar D}^2 H^M=0 $ \\
      &&&&\\
& & & $X^S$ & $ G^S $ \\
&&&&\\
& & & & $\Gamma^X$ \\
&&&&\\
\hline
\end{tabular}
\caption{Superspace version of table 1. The prepotentials are on the main diagonal, field strength superfields in the next and the Bianchi identities in the upper diagonal. Superfields in the same columns are of the same type. Starting on the left these are chiral, real, chiral spinor, real and chiral superfields. }
\end{table}

\subsection{Without Shifts}

We begin by reviewing how one embeds the usual potential fields in $N=1$ superspace using prepotential superfields \cite{Gates:1980ay} (see also \cite{Gates:1983nr}).  Following the superspace literature, we call these superfields ``prepotentials'' because there is another notion of superfields that deserve to be called potentials, namely we simply promote the bosonic $p$-forms to super $p$-forms, $\phi_{a_1\cdots a_p}\longrightarrow\Phi_{A_1\cdots A_p}$, where $A_i$ are superspace indices (e.g.\ running over $(x^a,\t^\al,\bar{\t}^{\dt\al})$).  After imposing certain constraints to ensure that the $\Phi_{[p]}$ give irreducible representations of supersymmetry, the potentials $\Phi_{[p]}$ can be solved in terms of the prepotentials we describe below \cite{Gates:1980ay,Gates:1983nr}.

\subsubsection{The Zero-Forms}

The zero form $a^A$ will be the real part of the bottom component of a chiral superfield $\Phi^A$, $\bar{D}_{\dt\al}\Phi^A=0$:
\be
a^A=\hlf\lp\Phi^A+\ov{\Phi}^A\rp\Big|.
\ee
In this section and below, the $|$ means that we should extract only the bottom component, i.e.\ set $\t=\bar{\t}=0$.
Gauge zero-forms differ from scalar fields in that they shift by a real constant under transformations $\delta \Phi^A = c^A$ (with $c^A\in \mathbb R$) leaving the classical action invariant.
The field strength invariant under this shift is\footnote{We apologize for the over-use of the letters $a$ and $F$, but it should hopefully be clear from context and indices whether we are talking about a bosonic field, a superfield, or an index.}
\be
\label{eq:NoShiftFA}
F^A=\frac{1}{2i}\lp\Phi^A-\ov{\Phi}^A\rp.
\ee
This field strength satisfies a Bianchi identity (the coefficients chosen will make more sense once we turn on the shifts)
\be
-\frac{1}{4}\bar{D}^2D_\al F^A=0.
\ee
To extract the component field strength, we take the $\theta\bar \theta$ component
\be
\label{eq:FSC0}
F^A_a=-\frac{1}{4}\lp\s_a\rp_{\al\dt\al}\ls D^\al,\bar{D}^{\dt\al}\rs F^A\Big|,
\ee
giving the bosonic field strength $F^A_a=\p_aa^A$.

Of course there are other component fields in the same multiplet, all of which are, like $a^A$, valued in $V_0$.  There is a real scalar partner to $a^A$, which we will call $\varphi^A$, given by
\be
\frac{1}{2i}\lp\Phi^A-\ov{\Phi}^A\rp\Big|.
\ee
Note that $\varphi^A$ is invariant under the shift above and therefore really a scalar instead of a zero-form.
There is also a complex auxiliary field
\be
-\frac{1}{4}D^2\Phi^A\Big|.
\ee
And finally there are the fermionic superpartners
\be
\psi^A_\al= 
\frac{1}{\sqrt{2}}D_\al\Phi^A\Big|
~~~\mathrm{and}~~~
\ov{\psi}^A_{\dot\al}= 
\frac{1}{\sqrt{2}}\bar{D}_{\dot\al}\ov{\Phi}^A\Big|.
\ee

\subsubsection{The One-Forms}

The vector $A^I_a$ naturally lives inside a real scalar superfield $V^I$, which suffers the gauge transformation,
\be
\d V^I=\frac{1}{2i}\lp\La^I-\ov{\La}^I\rp,
\ee
where $\La^I$ is chiral, $\bar{D}_{\dt\al}\La^I=0$.  The gauge field itself is extracted by
\be
A^I_a=\left. -\frac{1}{4}\lp\s_a\rp_{\al\dt\al}\ls D^\al,\bar{D}^{\dt\al}\rs V^I\right|,
\ee
and one can verify that
\be
\d A^I_a=\p_a\la^I,
\ee
where
\be
\la^I=\left.\hlf\lp\La^I+\ov{\La}^I\rp\right|.
\ee
Note that we can use the other components of $\La^I$ to go to Wess-Zumino gauge, in which we have (see e.g. \cite{Gates:1983nr, Wess:1992cp,Buchbinder:1998qv})
\be
V^I\Big|=D_\al V^I\Big|=\bar{D}_{\dot\al}V^I\Big|=D^2V^I\Big|=\bar{D}^2V^I\Big|=0.
\ee
The remaining component fields in $V^I$ consist of a real auxiliary field
\be
D^I=\frac{1}{16}\left\{D^2,\bar{D}^2\right\}V^I\Big|,
\ee
and fermions
\be
\la^I_\al=-\frac{i}{4}\bar{D}^2D_\al V^I\Big|,\qquad\ov{\la}^I_{\dot\al}=\frac{i}{4}D^2\bar{D}_{\dot\al}V^I\Big|.
\ee

The components $D^I$, $\la^I$, and $\ov{\la}^I$ are all gauge-invariant. We can make this manifest by constructing
an invariant field strength which is a chiral spinor superfield
\be
W^I_\al=-\frac{1}{4}\bar{D}^2D_\al V^I,
\ee
that contains (in addition to $D^I$ and $\la^I$) the appropriate component field strength
\be
\label{eq:FSC1}
F^I_{ab}=\left. -\frac{i}{2}\lp\lp\s_{ab}\rp_\al^{\hph{\al}\beta}D^\al W^I_\beta-\lp\bar{\s}_{ab}\rp^{\dt\al}_{\hph{\dt\al}\dt\beta}\bar{D}_{\dt\al}\ov{W}{}^{I\,\dt\beta}\rp\right|.
\ee
Furthermore, $W^I$ obeys the Bianchi identity
\be
\frac{1}{2i}\lp D^\al W^I_\al-\bar{D}_{\dt\al}\ov{W}{}^{I\,\dt\al}\rp=0.
\ee

\subsubsection{The Two-Forms}

The two-form potentials $B^M_{ab}$ reside in a chiral spinor superfield $\Sigma^M_\al$ in the same way that $F^I_{ab}$ lives inside of $W^I_\al$, i.e.\
\be
B^M_{ab}=\left. -\frac{i}{2}\lp\lp\s_{ab}\rp_\al^{\hph{\al}\beta}D^\al\Sigma^M_\beta-\lp\bar{\s}_{ab}\rp^{\dt\al}_{\hph{\dt\al}\dt\beta}\bar{D}_{\dt\al}\ov{\Sigma}{}^{M\,\dt\beta}\rp\right|.
\ee
The superfield $\Sigma^M_\al$ has a gauge transformation
\be
\d\Sigma^M_\al=-\frac{1}{4}\bar{D}^2D_\al U^M,
\ee
where $U^M$ is a real scalar superfield.  We of course have
\be
\La^M_a=\left. -\frac{1}{4}\lp\s_a\rp_{\al\dt\al}\ls D^\al,\bar{D}^{\dt\al}\rs U^M\right|,
\ee
and
\be
\d B^M_{ab}=2\p_{[a}\La^M_{b]}.
\ee
The remaining components of $U^M$ either drop out entirely (if they are part of a chiral superfield plus its conjugate), or they can be used to set some components of $\Sigma^M_\al$ to zero, in an analog of Wess-Zumino gauge.  Explicitly, we can set
\be
\Sigma^M_\al\Big|=\ov{\Sigma}^M_{\dot\al}\Big|=0,
\ee
and we can set the real part of $D^\al\Sigma^M_\al$ (which also equals the real part of $\bar{D}_{\dot\al}\ov{\Sigma}{}^{M\,\dot\al}$) to zero.  The remaining gauge-invariant components are a real scalar
\be
\ell^M=\frac{1}{4i}\lp D^\al\Sigma^M_\al-\bar{D}_{\dot\al}\ov{\Sigma}{}^{M\,\dot\al}\rp\Big|,
\ee
and fermions
\be
\chi^M_\al=-\frac{1}{4\sqrt{2}}D^2\Sigma^M_\al\Big|,\qquad\ov{\chi}_{\dot\al}=-\frac{1}{4\sqrt{2}}\bar{D}^2\ov{\Sigma}^M_{\dot\al}\Big|.
\ee
The corresponding invariant field strength is
\be
H^M=\frac{1}{2i}\lp D^\al\Sigma^M_\al-\bar{D}_{\dt\al}\ov{\Sigma}{}^{M\,\dt\al}\rp,
\ee
with
\be
\label{eq:FSC2}
F^M_{abc}=\left.\frac{1}{8}\e_{abcd}\s^d_{\al\dt\al}\ls D^\al,\bar{D}^{\dt\al}\rs H^M\right|.
\ee
This invariant superfield strength obeys the Bianchi identity
\be
-\frac{1}{4}\bar{D}^2H^M=0.
\ee

\subsubsection{The Three-Forms}

The three-form $C^S_{abc}$ is embedded in a real scalar superfield $X^S$,
\be
C^S_{abc}=\left.\frac{1}{8}\e_{abcd}\s^d_{\al\dt\al}\ls D^\al,\bar{D}^{\dt\al}\rs X^S\right|.
\ee
The gauge transformation is parameterized by a chiral spinor superfield $\Upsilon^S_\al$, with
\be
\La^S_{ab}=\left. -\frac{i}{2}\lp\lp\s_{ab}\rp_\al^{\hph{\al}\beta}D^\al\Upsilon^S_\beta-\lp\bar{\s}_{ab}\rp^{\dt\al}_{\hph{\dt\al}\dt\beta}\bar{D}_{\dt\al}\ov{\Upsilon}{}^{S\,\dt\beta}\rp\right|,
\ee
and the superfield transformation is
\be
\d X^S=\frac{1}{2i}\lp D^\al\Upsilon_\al-\bar{D}_{\dt\al}\ov{\Upsilon}{}^{\dt\al}\rp.
\ee
Going to (an analog of) Wess-Zumino gauge, we can ensure that
\be
X^S\Big|=D_\al X^S\Big|=\bar{D}_{\dot\al}X^S\Big|=0,
\ee
leaving us with a complex scalar
\be
y^S=-\frac{1}{4}\bar{D}^2X^S\Big|,
\ee
a real auxiliary scalar,
\be
z^S=\frac{1}{32}\left\{D^2,\bar{D}^2\right\}X^S\Big|,
\ee
and fermions
\be
\eta^S_\al=-\frac{1}{4\sqrt{2}}\bar{D}^2D_\al X^S\Big|,\qquad\ov{\eta}^S_{\dot\al}=-\frac{1}{4\sqrt{2}}D^2\bar{D}_{\dot\al}X^S\Big|.
\ee

The field strength is a chiral superfield,
\be
\label{eq:FSC3}
G^S=-\frac{1}{4}\bar{D}^2X^S,\qquad
F^S_{abcd}=\left.\frac{i}{8}\e_{abcd}\lp D^2G^S-\bar{D}^2\ov{G}{}^S\rp\right|.
\ee
There's no corresponding Bianchi identity
since the bosonic field strength $F^S_{abcd}$ is automatically closed by virtue of being a 4-form.

\subsubsection{The Four-Forms}

Finally, the four-form potential $D^X_{abcd}$ can be placed in a chiral superfield $\G^X$,
\be
D^X_{abcd}=\left.\frac{i}{8}\e_{abcd}\lp D^2\G^X-\bar{D}^2\ov{\G}{}^X\rp\right|.
\ee
The gauge parameter lives in a real scalar superfield $\Xi^X$,
\be
\La^X_{abc}=\left.\frac{1}{8}\e_{abcd}\s^d_{\al\dt\al}\ls D^\al,\bar{D}^{\dt\al}\rs\Xi^X\right|,
\ee
and the superfield transforms as
\be
\d\G^X=-\frac{1}{4}\bar{D}^2\Xi^X.
\ee
There is no field strength in this case, and the space of gauge transformations is large enough to gauge away every component of $\G^X$ except for $D^X_{abcd}$ (and even this can be gauged away locally, using the residual bosonic symmetry parameterized by $\La^X_{abc}$).

\subsection{With Shifts}

With the details above, it is not hard to incorporate the shifts.  For instance, the zero-form now transforms, so we should have (we drop the $^{(p)}$ superscripts on $q$ since the degree is clear from the indices),
\be
\d\Phi^A=\lp q\cdot\La\rp^A,
\ee
and correspondingly we must deform the field strength (\ref{eq:NoShiftFA}) to,
\be
F^A=\frac{1}{2i}\lp\Phi^A-\ov{\Phi}^A\rp-\lp q\cdot V\rp^A.
\ee
This modifies the Bianchi identity to
\be
-\frac{1}{4}\bar{D}^2D_\al F^A=-\lp q\cdot W_\al\rp^A.
\ee

Proceeding similarly for the other fields, we arrive at the variations\footnote{\label{fm:4FormGaugeParam}There is one more possibility, which is that we could add a term $(q^{(4)})^X_m\Omega^m$
(with $m$ indexing the space $V_5$)
to the last line of (\ref{E:superGaugeXf}), where $\Omega^m$ is a chiral superfield.  In components, this would generate a shift $\d D^X_{abcd}=q^X_m\La^m_{abcd}$, but there is no corresponding field labeled by $m$ for which $\La^m$ is an ordinary gauge parameter.  In the dimensional reduction case, this would happen only if $q>d$, i.e.\ we are reducing a form in $D$ dimensions whose degree is greater than the spacetime dimension $d$.}
\begin{align}
\label{E:superGaugeXf}
\begin{array}{lclccl}
&\delta \Phi^A &=&  &+ & \lp q\cdot\Lambda\rp^A
\cr
&\delta V^I &= &\frac{1}{2i}\lp\La^I-\ov{\La}^I\rp & + & \lp q\cdot U\rp^I
\cr
&\delta \Sigma^M_\alpha &=& -\frac14 \bar D^2D_\alpha U^M &+ & \lp q\cdot\Upsilon_\alpha\rp^M
\cr
&\delta X^S &=& \frac{1}{2i} \left( D^\alpha \Upsilon^S_\alpha - \bar D_{\dt \alpha}\ov{\Upsilon}{}^{S\,\dt \alpha}\right)&+ & \lp q\cdot\Xi\rp^S,\cr
&\delta\G^X &=& -\frac{1}{4}\bar{D}^2\Xi^X.&&
\end{array}
\end{align}
These prompt us to construct invariant field strength superfields
\begin{align}
\label{eq:SuperfieldStengths}
\begin{array}{lclccl}
&E^Z &=&
	&-& \lp q\cdot\Phi\rp^Z
\cr
&F^A &= &\frac 1{2i} \left( \Phi^A -\bar \Phi^A\right)
	& - & \lp q\cdot V\rp^A
\cr
&W^I_\alpha &=& -\frac14 \bar D^2D_\alpha V^I
	&- & \lp q\cdot\Sigma_\alpha\rp^I
\cr
&H^M &=& \frac 1{2i} \left( D^\alpha \Sigma^M_\alpha - \bar D_{\dt \alpha}\bar \Sigma^{M\dt \alpha}\right)
	& -& \lp q\cdot X\rp^M
\cr
&G^S &=& -\frac14 \bar D^2 X^S&- & \lp q\cdot\G\rp^S .
\end{array}
\end{align}
Notice that we have also introduced the ``zero-form field strength'' $E^Z$, which is a chiral superfield, with component
\be
F^Z=\left.\hlf\lp E^Z+\ov{E}^Z\rp\right|.
\ee
Finally, these field strengths obey Bianchi identities
\begin{align}
\label{E:superBI}
\begin{array}{lclccl}
&0 &=& \frac1{2i} \left( E^Z - \bar E^Z \right)
	&+& \lp q\cdot F\rp^Z
\cr
&0 &=& -\frac14 \bar D^2D_\alpha F^A
	&+& \lp q\cdot W_\alpha\rp^A
\cr
&0 &=& \frac 1{2i} \left( D^\alpha W^I_\alpha - \bar D_{\dt \alpha}\ov{W}{}^{I\, \dt \alpha}\right)
 	&+& \lp q\cdot H\rp^I
\cr
&0 &=& -\frac14 \bar D^2 H^M &+& \lp q\cdot G\rp^M.
\end{array}
\end{align}
Note the beautiful symmetry\footnote{If we have one more map, $(q^{(-2)})^m_Z$, then we could make the symmetry even clearer by adding a line $0=q^m_ZE^Z$ at the top of the third set of equations, (\ref{E:superBI}).  Indeed, in the dimensional reduction example where $q$ is just the exterior derivative on the internal space, we do have such a map; $q^{(-2)}$ is just the exterior derivative acting on $(q+2)$-forms.  For the other possible lack of symmetry, see footnote \ref{fm:4FormGaugeParam}.} between (\ref{E:superGaugeXf}), (\ref{eq:SuperfieldStengths}) and (\ref{E:superBI}). The same operations appear in each set of equations to relate forms of different space-time degree.

The rest of the discussion goes mostly the same.  We can still access Wess-Zumino gauge\footnote{Actually, this depends a bit delicately on the fact that $q\cdot q=0$.  For example, suppose we do an arbitrary $U^M$ transformation.  This will not generally leave $V^I$ in Wess-Zumino gauge, so we need to perform a compensating $\La^I(U^M)$ transformation to return $V^I$ to Wess-Zumino gauge.  A priori, this compensating transformation would affect the scalars, but in fact they remain invariant provided $q^{(0)}\cdot q^{(1)}=0$.} for $V^I$, $\Sigma^M$, $X^S$, and $\G^X$, and these fields still have the same component expansions.  The field strengths have been modified, but the relations to components (\ref{eq:FSC0}), (\ref{eq:FSC1}), (\ref{eq:FSC2}), and (\ref{eq:FSC3}) are the same, only now in terms of the properly gauge-invariant bosonic field strengths (\ref{eq:BosonicFieldStrengths}).

\subsection{Gauge Invariant Kinetic Terms}

Since the superfield strengths are gauge invariant, a supersymmetric and gauge invariant Lagrangian can be obtained by combining superfield strengths into chiral superfields and integrating them over half of superspace or into real combinations and integrating over all of superspace.
Here, we present the simplest possibility, namely that we have a constant metric on each $V_p$ and use it to build simple quadratic combinations of the field strengths.  Explicitly,
\begin{equation}
\begin{split}
\int d^4xd^4\t\, g_{AB}F^AF^B= & \int d^4x\,g_{AB}\Big[ -\hlf F^{A\,a}F^B_a-\hlf\p^a\varphi^A\p_a\varphi^B+\hlf f^A\ov{f}^B-\frac{i}{2}\psi^A\s^a\p_a\ov{\psi}^B
\\
&
 -q^B_I\lp\varphi^AD^I +\frac{1}{\sqrt{2}}\psi^A\la^I+\frac{1}{\sqrt{2}}\ov{\psi}^A\ov{\la}^I\rp\Big],
\end{split}
\end{equation}
\begin{equation}
\begin{split}
\operatorname{Re}\left( \int d^4xd^2\t\,g_{IJ}W^IW^J\right)= & \int d^4x\,\Big[\operatorname{Im}(g_{IJ})\lp -2D^Iq^J_M\ell^M+\frac{1}{4}\e^{abcd}F^I_{ab}F^J_{cd}\rp \\
& +\operatorname{Re}(g_{IJ})\lp -\hlf F^{I\,ab}F^J_{ab}+D^ID^J-q^I_Mq^J_N\ell^M\ell^N-2i\la^I\s^a\p_a\ov{\la}^J\rp \\
& +\sqrt{2}ig_{IJ}q^J_M\la^I\chi^M-\sqrt{2}i\ov{g_{IJ}}q^J_M\ov{\la}^I\ov{\chi}^M\Big],
\end{split}
\end{equation}
\begin{equation}
\begin{split}
\int d^4xd^4\t\,g_{MN}H^MH^N= & \int d^4x\,g_{MN}\Big[ \frac{1}{3}F^{M\,abc}F^N_{abc}+2\p^a\ell^M\p_a\ell^N-2i\chi^M\s^a\p_a\ov{\chi}^N\\
& +2q^N_S\lp -\ell^Mz^S-i\chi^M\eta^S+i\ov{\chi}^M\ov{\eta}^S\rp+2q^M_Sq^N_Ty^S\ov{y}^T\Big],
\end{split}
\end{equation}
and\footnote{On dimensional grounds, we need to take this $D$-term action to give a kinetic term for $C^S_{abc}$, rather than the $F$-term possibility $\int d^2\t g_{ST}G^SG^T$.}
\begin{equation}
\int d^4xd^4\t\,g_{ST}G^S\ov{G}^T
=  \int d^4x\,g_{ST}\Big[-\frac{1}{24}F^{S\,abcd}F^T_{abcd}+z^Sz^T-\p^ay^S\p_a\ov{y}^T -i\eta^S\s^a\p_a\ov{\eta}^T\Big].
\end{equation}
Here $g_{AB}$, $g_{MN}$, and $g_{ST}$ are constant real metrics.  $g_{IJ}$ can a priori be complex, and unlike in the usual case (without shifts), the action proportional to the imaginary part of $g_{IJ}$ is not purely topological.

\section{Bosonic Chern-Simons Actions}

With the invariant field strengths constructed in section \ref{sec:Superfields}, it is easy to write down gauge-invariant supersymmetric actions simply by building real scalar (or chiral) combinations and integrating them over all (or half) of superspace.  However, there is another important  possibility, which is to have a Lagrangian that is not gauge invariant, but whose variation vanishes when integrated over superspace. This is the hallmark of a Chern-Simons form. In the next subsection we will review the typical example of this in the bosonic case, where we build a $d$-form in $d$ dimensions by wedging one potential $\phi_{[p_0]}$ and some number of field strengths $F_{[p_1]},\cdots,F_{[p_n]}$, with $\sum_{i=0}^np_i=d$.  Without shifts this would be gauge invariant when integrated, since its variation is an exact form.  This is what we will mean when we say ``Chern-Simons actions''.  With shifts, we still have a chance of building something invariant by taking linear combinations of such terms.  After explaining the bosonic case in this section, we will construct the supersymmetric analog in the next section.

\subsection{Actions}

Again, we restrict to the case $d=4$, and denote our potential $p$-form fields $a^A$, $A^I$, $B^M$, $C^S$, and $D^X$, for $p$ running from zero to four respectively.  We will consider the cases $n=0,1,2$, where $n$ is the number of field strengths.  It is not difficult to work out the story for higher $n$, though such actions are then higher order than quadratic in derivatives.

\subsubsection{Linear Chern-Simons Terms}
\label{S:CS0}
For $n=0$, we can only construct a four-form by using $D^X$,
\be
S_{0,CS}=\int\al_XD^X,
\ee
where $\al_X$ are some set of constants.  These terms are gauge invariant for any choice\footnote{This is true up to the possible caveat mentioned in footnote \ref{fm:4FormGaugeParam}.  In this case, gauge invariance under the shifts parameterized by $\Om^m$ would require $\al_Xq^X_m=0$.} of $\al_X$, since $\d D^X=d\La^X$ is exact.
An example of this sort of coupling is given by D3-branes, on which we have a coupling $\int_{D3} C_{[4]}$.

\subsubsection{Quadratic Chern-Simons Terms}

For $n=1$, we have five possible terms,
\be
\label{eq:S1CS}
S_{1,CS}=\int\left\{\al_{1AS}a^AF^S_{[4]}+\al_{2IM}A^I\w F^M_{[3]}+\al_{3MI}B^M\w F^I_{[2]}+\al_{4SA}C^S\w F^A_{[1]}+\al_{5XZ}D^XF^Z_{[0]}\right\}.
\ee
The $BF$ coupling proportional to $\al_3$ is probably the most familiar of these terms, but they can all occur.  Note also that in the case without shifts the terms are not all independent: The $\al_1$ and $\al_4$ terms are related to each other by integration by parts, as are the $\al_2$ and $\al_3$ terms.  With shifts this is no longer true (although there can still be relations).

Under the gauge transformations (\ref{e4}), we have
\begin{align}
\d S_{1,CS}=& \int\left\{\al_{1AS}q^A_I\La^I_{[0]}F^S_{[4]}+\al_{2IM}\lp d\La^I_{[0]}+q^I_N\La^N_{[1]}\rp\w F^M_{[3]}+\al_{3MI}\lp d\La^M_{[1]}+q^M_S\La^S_{[2]}\rp\w F^I_{[2]}\right.\non\\
& \qquad\left.+\al_{4SA}\lp d\La^S_{[2]}+q^S_X\La^X_{[3]}\rp\w F^A_{[1]}+\al_{5XZ}d\La^X_{[3]}F^Z_{[0]}\right\}\non\\
=& \int\left\{\lp\al_{1AS}q^A_I+\al_{2IM}q^M_S\rp\La^I_{[0]}F^S_{[4]}+\lp\al_{2IN}q^I_M-\al_{3MI}q^I_N\rp\La^M_{[1]}\w F^N_{[3]}\right.\non\\
& \qquad\left. +\lp\al_{3MI}q^M_S+\al_{4SA}q^A_I\rp\La^S_{[2]}\w F^I_{[2]}+\lp\al_{4SA}q^S_X-\al_{5XZ}q^Z_A\rp\La^X_{[3]}\w F^A_{[1]}\right\},
\end{align}
where we have integrated by parts and used the Bianchi identities for the field strengths.  In order for this to be gauge invariant, we must require each of the combinations in parentheses to vanish, i.e.\
\bea
0 &=& \al_{1AS}q^A_I+\al_{2IM}q^M_S,\non\\
0 &=& \al_{2IN}q^I_M-\al_{3MI}q^I_N,\non\\
\label{eq:QuadraticCSVariation}
0 &=& \al_{3MI}q^M_S+\al_{4SA}q^A_I,\\
0 &=& \al_{4SA}q^S_X-\al_{5XZ}q^Z_A.\non
\eea

\subsubsection{Cubic Chern-Simons Terms}

Now we have nine possible terms
\begin{align}
\label{eq:S2CS}
S_{2,CS}=& \int\left\{\vphantom{F^{Z'}_{[0]}}\al_{1AZS}a^AF^Z_{[0]}F^S_{[4]}+\al_{2ABM}a^AF^B_{[1]}\w F^M_{[3]}+\al_{3AIJ}a^AF^I_{[2]}\w F^J_{[2]}+\al_{4IZM}A^I\w F^Z_{[0]}F^M_{[3]}\right.\non\\
& \qquad\left.+\al_{5IAJ}A^I\w F^A_{[1]}\w F^J_{[2]}+\al_{6MZI}B^M\w F^Z_{[0]}F^I_{[2]}+\al_{7MAB}B^M\w F^A_{[1]}\w F^B_{[1]}\right.\non\\
& \qquad\left. +\al_{8SZA}C^S\w F^Z_{[0]}F^A_{[1]}+\al_{9XZZ'}D^XF^Z_{[0]}F^{Z'}_{[0]}\right\}.
\end{align}
Without loss of generality we can take $\al_{3AIJ}=\al_{3AJI}$ and $\al_{9XZZ'}=\al_{9XZ'Z}$ to be symmetric in their last two indices, and $\al_{7MAB}=-\al_{7MBA}$ to be antisymmetric.  The $\al_3$ term is the familiar axionic coupling in four dimensions.  The variation is given, after integration by parts and use of Bianchi identities, by
\begin{align}
\d S_{2,CS}=& \int\left\{\vphantom{\lp q^{Z'}_A\rp}\La^I_{[0]}\ls\lp\al_{1AZS}q^A_I+\al_{4IZM}q^M_S\rp F^Z_{[0]}F^S_{[4]}+\lp\al_{2BAM}q^B_I+\al_{4IZM}q^Z_A-\al_{5IAJ}q^J_M\rp F^A_{[1]}\w F^M_{[3]}\right.\right.\non\\
& \qquad\left.\left. +\lp\al_{3AJK}q^A_I+\al_{5IAJ}q^A_K\rp F^J_{[2]}\w F^K_{[2]}\rs+\La^M_{[1]}\w\ls\lp\al_{4IZN}q^I_M-\al_{6MZI}q^I_N\rp F^Z_{[0]}F^N_{[3]}\right.\right.\non\\
& \qquad\left.\left.+\lp\al_{5JAI}q^J_M-\al_{6MZI}q^Z_A+2\al_{7MAB}q^B_I\rp F^A_{[1]}\w F^I_{[2]}\rs\right.\non\\
& \quad\left.+\La^S_{[2]}\w\ls\lp\al_{6MZI}q^M_S+\al_{8SZA}q^A_I\rp F^Z_{[0]}F^I_{[2]}+\lp\al_{7MAB}q^M_S-\al_{8SZA}q^Z_B\rp F^A_{[1]}\w F^B_{[1]}\rs\right.\non\\
& \quad\left. +\La^X_{[3]}\w\lp\al_{8SZA}q^S_X-2\al_{9XZZ'}q^{Z'}_A\rp F^Z_{[0]}F^A_{[1]}\right\}.
\end{align}

Recalling that $F^Z_{[0]}=-q^Z_Aa^A$ always carries a $q^Z_A$, the vanishing of this variation is equivalent to four equations that are linear in the $q$s,
\bea
\label{eq:CubicInv1}
0 &=& \al_{2BAM}q^B_I+\al_{4IZM}q^Z_A-\al_{5IAJ}q^J_M,\non\\
0 &=& \al_{3AJK}q^A_I+\al_{5IA(J}q^A_{K)},\non\\
0 &=& \al_{5JAI}q^J_M-\al_{6MZI}q^Z_A+2\al_{7MAB}q^B_I,\\
0 &=& \al_{7MAB}q^M_S-\al_{8SZ[A}q^Z_{B]},\non
\eea
and four that have an extra factor of $q^Z_A$,
\bea
\label{eq:CubicInv2}
0 &=& \al_{1BZS}q^B_Iq^Z_A+\al_{4IZM}q^M_Sq^Z_A,\non\\
0 &=& \al_{4IZN}q^I_Mq^Z_A-\al_{6MZI}q^I_Nq^Z_A,\non\\
0 &=& \al_{6MZI}q^M_Sq^Z_A+\al_{8SZB}q^B_Iq^Z_A,\\
0 &=& \al_{8SZB}q^S_Xq^Z_A-2\al_{9XZZ'}q^Z_Aq^{Z'}_B.\non
\eea

\subsection{Descent Formalism}

Each of the cases above (linear, quadratic, and cubic) can be combined into a nicely packaged formalism by writing
\be
\label{eq:BosonicDescentAction}
S_{CS}=\int\left\{a^Ac_{[4]A}+A^I\w c_{[3]I}+B^M\w c_{[2]M}+C^S\w c_{[1]S}+D^Xc_{[0]X}\right\},
\ee
where each $c_{[4-p]I_p}$ is a polynomial in the field strengths.  This action is invariant if
\be
\label{eq:BosonicDescentEq}
q^{I_p}_{I_{p+1}}c_{[4-p]I_p}-\lp -1\rp^pdc_{[3-p]I_{p+1}}=0,
\ee
for each $p=0,\cdots,3$.

In this formalism, the linear case is given by the solution $c_{[1]}=c_{[2]}=c_{[3]}=c_{[4]}=0$, $c_{[0]X}=\al_X$ is constant.
The quadratic case has
\bea
c_{[4]A} &=& \al_{1AS}F^S_{[4]},\non\\
c_{[3]I} &=& \al_{2IM}F^M_{[3]},\non\\
c_{[2]M} &=& \al_{3MI}F^I_{[2]},\\
c_{[1]S} &=& \al_{4SA}F^A_{[1]},\non\\
c_{[0]X} &=& \al_{5XZ}F^Z_{[0]}.\non
\eea
Note that the requirement (\ref{eq:BosonicDescentEq}) that the forms $c_{[p]}$ must satisfy is very similar to the Bianchi identities (\ref{eq:BosonicBianchis}), except that we replace $q$ by its transpose.  For the quadratic case in particular, the requirements derived from (\ref{eq:QuadraticCSVariation}) are equivalent to the statement that the $\al_i$ give a pairing on the complex $V_\newbullet$ with respect to which the adjoint of $q$ is just the transpose of $q$.  Then the descent relations (\ref{eq:BosonicDescentEq}) simply follow from the Bianchi identities (\ref{eq:BosonicBianchis}).

Finally, for the cubic case, we read off
\bea
\label{eq:BosonicCubicDescent}
c_{[4]A} &=& \al_{1AZS}F^Z_{[0]}F^S_{[4]}+\al_{2ABM}F^B_{[1]}\w F^M_{[3]}+\al_{3AIJ}F^I_{[2]}\w F^J_{[2]},\non\\
c_{[3]I} &=& \al_{4IZM}F^Z_{[0]}F^M_{[3]}+\al_{5IAJ}F^A_{[1]}\w F^J_{[2]},\non\\
c_{[2]M} &=& \al_{6MZI}F^Z_{[0]}F^I_{[2]}+\al_{7MAB}F^A_{[1]}\w F^B_{[1]},\\
c_{[1]S} &=& \al_{8SZA}F^Z_{[0]}F^A_{[1]},\non\\
c_{[0]X} &=& \al_{9XZZ'}F^Z_{[0]}F^{Z'}_{[0]}.\non
\eea

\subsection{Examples from Dimensional Reduction}

\subsubsection{Dimensional Reduction from 5 to 4}

Consider a theory in five dimensions with a vector $\widetilde{A}$. It is easy to generalize this story to multiple five-dimensional vectors.  This theory can have a Chern-Simons coupling of the form
\be
S_{5D,CS}=\g\int \widetilde{A}\w\widetilde{F}\w\widetilde{F},
\ee
where $\g$ is a constant.  Upon reduction on a circle (with coordinate $y$ and radius $R$), the five-dimensional vector gives rise to an infinite set (the KK tower) of axionic scalars $a^{(y)}(x)=\widetilde{A}_y(x,y)$ and an infinite set of four-dimensional vectors $A_a^{(y)}(x)=\widetilde{A}_a(x,y)$.  We also have a ``matrix''
\be
\lp q^{(0)}\rp^{(y)}_{(y')}=\frac{\p}{\p y}\d(y-y'),
\ee
and gauge transformation and field strengths
\be
\d a^{(y)}=\frac{\p}{\p y}\La^{(y)}_{[0]},\qquad F^{(y)}_{[1]}=da^{(y)}-\frac{\p}{\p y}A^{(y)},
~~~\mathrm{and}~~~
F^{(y)}_{[2]}=dA^{(y)} .
\ee

In terms of four-dimensional couplings, the five-dimensional Chern-Simons action would now be written as
\be
S_{5D,CS}=\int\ls\al_{3(y)(y')(y'')}a^{(y)}F^{(y')}_{[2]}\w F^{(y'')}_{[2]}+\al_{5(y)(y')(y'')}A^{(y)}\w F^{(y')}_{[1]}\w F^{(y'')}_{[2]}\rs,
\ee
where
\be
\al_{3(y)(y')(y'')}=\g\d(y-y')\d(y-y'')
~~~\mathrm{and}~~~
\al_{5(y)(y')(y'')}=2\g\d(y-y')\d(y-y'').
\ee

To compare with more traditional presentations of Ka{\l}u\.za-Klein theory, let us do a Fourier expansion,
\be
\widetilde{A}_4(x,y)=\sum_{n\in\Z}a^n(x)e^{iny/R},\qquad\widetilde{A}_a(x,y)=\sum_{N\in\Z}A^N_a(x)e^{iNy/R},
\ee
with reality conditions $(a^n)^\ast=a^{-n}$, $(A^N_a)^\ast=A^{-N}_a$.  We used different labels $n$ and $N$ to emphasize that these label bases for the space $V_0$ and $V_1$ respectively.  Similarly, for the gauge parameter we have an expression
\be
\La(x,y)=\sum_{N\in\Z}\La^Ne^{iNy/R}.
\ee

In this basis,
\be
\label{E:FourierGauge}
\d a^n=\frac{in}{R}\d^n_N\La^N,\qquad\d A^N=d\La^N,
\ee
\be
F^n_{[1]}=da^n-\frac{in}{R}\d^n_NA^N,\qquad F^N_{[2]}=dA^N,
\ee
and
\be
\al_{3nMP}=\g R\d_{n+M+P,0},\qquad\al_{5NmP}=2\g R\d_{N+m+P,0}.
\ee
Then one can verify that the action
\be
S_{5D,CS}=\int\ls\sum_{n,M,P}\al_{3nMP}a^nF^M_{[2]}\w F^P_{[2]}+\sum_{N,m,P}\al_{5NmP}A^N\w F^m_{[1]}\w F^P_{[2]}\rs
\ee
is invariant. 

\subsubsection{Dimensional Reduction from 11 to 4}

Eleven-dimensional supergravity has a three-form potential $C_{MNP}$.  Upon reduction to four dimensions, this gives us potentials
\bea
a_{(ijk;y)}(x) &=& C_{ijk}(x,y),\non\\
\lp A_{(ij;y)}\rp_a(x) &=& C_{aij}(x,y),\non\\
\lp B_{(i;y)}\rp_{ab}(x) &=& C_{abi}(x,y),\\
\lp C_{(;y)}\rp_{abc}(x) &=& C_{abc}(x,y).\non
\eea
Note that there is no four-form $D^X$.

The matrices $q$ are given by (\ref{eq:DimRedq}) with $n=3$.  The corresponding field strengths are
\bea
F_{(ijk\ell;y)}(x) &=& 4\p_{[i}C_{jk\ell]}(x,y),\non\\
\lp F_{(ijk;y)}\rp_a(x) &=& \p_aC_{ijk}(x,y)-3\p_{[i}C_{|a|jk}(x,y),\non\\
\lp F_{(ij;y)}\rp_{ab}(x) &=& 2\p_{[a}C_{b]ij}(x,y)+2\p_{[i}C_{|ab|j]}(x,y),\\
\lp F_{(i;y)}\rp_{abc}(x) &=& 3\p_{[a}C_{bc]i}(x,y)-\p_iC_{abc}(x,y),\non\\
\lp F_{(;y)}\rp_{abcd}(x) &=& 4\p_{[a}C_{bcd]}(x,y).\non
\eea
These satisfy Bianchi identities (\ref{eq:BosonicBianchis}) in the form,
\bea
0 &=& -5\p_{[i}F_{[0](jk\ell m];y)},\non\\
dF_{[0](ijk\ell;y)} &=& 4\p_{[i}F_{[1](jk\ell];y)},\non\\
dF_{[1](ijk;y)} &=& -3\p_{[i}F_{[2](jk];y)},\non\\
dF_{[2](ij;y)} &=& 2\p_{[i}F_{[3](j];y)},\\
dF_{[3](i;y)} &=& -\p_iF_{[4](;y)},\non\\
dF_{[4](;y)} &=& 0.\non
\eea

The eleven-dimensional theory has a Chern-Simons term
\be
\k\int C\w dC\w dC,
\ee
where $\k$ is a constant.  Reducing to four dimensions we can write it in the form (\ref{eq:BosonicDescentAction}), with
\bea
\label{eq:Bosonic11Descent}
c_{[4]}^{(ijk;y)} &=& \frac{\k}{3!4!}\e^{ijk\ell mnp}\lp 2F_{[0](\ell mnp;y)}F_{[4](;y)}-8F_{[1](\ell mn;y)}\w F_{[3](p;y)}+6F_{[2](\ell m;y)}\w F_{[2](np;y)}\rp,\non\\
c_{[3]}^{(ij;y)} &=& \frac{\k}{2!5!}\e^{ijk\ell mnp}\lp 10F_{[0](k\ell mn;y)}F_{[3](p;y)}+20F_{[1](k\ell m;y)}\w F_{[2](np;y)}\rp,\non\\
\label{eq:Bosonic11Descent}
c_{[2]}^{(i;y)} &=& \frac{\k}{6!}\e^{ijk\ell mnp}\lp 30F_{[0](jk\ell m;y)}F_{[2](np;y)}-20F_{[1](jk\ell;y)}\w F_{[1](mnp;y)}\rp,\\
c_{[1]}^{(;y)} &=& \frac{70\k}{7!}\e^{ijk\ell mnp}F_{[0](ijk\ell;y)}F_{[1](mnp;y)}.\non
\eea
We can verify that these satisfy (\ref{eq:BosonicDescentEq}).

We can also read off the $\al$ coefficients by comparing (\ref{eq:Bosonic11Descent}) with (\ref{eq:BosonicCubicDescent}).  The result is
\bea
\label{eq:11Dalphas}
\al_1^{(ijk;y)(\ell mnp;y')(;y'')} &=& \frac{\k}{72}\e^{ijk\ell mnp}\d(y-y')\d(y-y''),\non\\
\al_2^{(ijk;y)(\ell mn;y')(p;y'')} &=& -\frac{\k}{18}\e^{ijk\ell mnp}\d(y-y')\d(y-y''),\non\\
\al_3^{(ijk;y)(\ell m;y')(np;y'')} &=& \frac{\k}{24}\e^{ijk\ell mnp}\d(y-y')\d(y-y''),\non\\
\al_4^{(ij;y)(k\ell mn;y')(p;y'')} &=& \frac{\k}{24}\e^{ijk\ell mnp}\d(y-y')\d(y-y''),\non\\
\al_5^{(ij;y)(k\ell m;y')(np;y'')} &=& \frac{\k}{12}\e^{ijk\ell mnp}\d(y-y')\d(y-y''),\\
\al_6^{(i;y)(jk\ell m;y')(np;y'')} &=& \frac{\k}{24}\e^{ijk\ell mnp}\d(y-y')\d(y-y''),\non\\
\al_7^{(i;y)(jk\ell;y')(mnp;y'')} &=& -\frac{\k}{36}\e^{ijk\ell mnp}\d(y-y')\d(y-y''),\non\\
\al_8^{(;y)(ijk\ell;y')(mnp;y'')} &=& \frac{\k}{72}\e^{ijk\ell mnp}\d(y-y')\d(y-y'').\non
\eea
There is no $\al_9$ because there is no four-form potential.

\section{Superfield Chern-Simons Actions}

Now we make use of the superfields we defined in section \ref{sec:Superfields} and write down supersymmetrizations of these Chern-Simons actions.

\subsection{Actions}

\subsubsection{Linear Chern-Simons Terms}

In the case of the linear Chern-Simons term (cf.~\S~\ref{S:CS0}), it turns out that, surprisingly, the bosonic action is already supersymmetric, since we have
\be
S_{0,SCS}=\operatorname{Re}\ls i\int d^4xd^2\t\al_X\G^X\rs=\int\al_XD^X=S_{0,CS}.
\ee

As before it is gauge invariant,
\be
\d S_{0,SCS}=\operatorname{Re}\ls i\int d^4xd^2\t\al_X\lp -\frac{1}{4}\bar{D}^2\Xi^X\rp\rs=\operatorname{Re}\ls i\int d^4xd^4\t\al_X\Xi^X\rs=0,
\ee
where in the last step we used that $d^4\t$, $\al_X$, and $\Xi^X$ are real, so the quantity in square brackets is purely imaginary. Note that this Fayet-Iliopulos type term is proportional to the $F$-term of the chiral multiplet $\Gamma^X$ and may play an interesting role in the breaking of supersymmetry.

\subsubsection{Quadratic Chern-Simons Terms}

In this case, the supersymmetrization of the Chern-Simons action has the form
\begin{align}
S_{1,SCS}=& \int d^4xd^4\t\lp\al_{2IM}V^IH^M-\al_{4SA}X^SF^A\rp\non\\
& +\operatorname{Re}\ls i\int d^4xd^2\t\lp\al_{1AS}\Phi^AG^S+\al_{3MI}\Sigma^{M\,\al}W^I_\al+\al_{5XZ}\G^XE^Z\rp\rs.
\end{align}
When expanded into components, the resulting action contains (\ref{eq:S1CS}),
but will have many other pieces involving
the superpartners as well as additional bosons required by supersymmetry.

Under the supersymmetric gauge transformations (\ref{E:superGaugeXf}), the action changes by
\begin{align}
\d S_{1,SCS}=& \int d^4xd^4\t\lp\al_{2IM}\lp\frac{1}{2i}\lp\La^I-\ov{\La}^I\rp+\lp q\cdot U\rp^I\rp H^M\right.\non\\
& \qquad\left. -\al_{4SA}\lp\frac{1}{2i}\lp D^\al\Upsilon^S_\al-\bar{D}_{\dot\al}\ov{\Upsilon}^{S\,\dot\al}\rp+\lp q\cdot\Xi\rp^S\rp F^A\rp\non\\
& \quad+\operatorname{Re}\ls i\int d^4xd^2\t\lp\al_{1AS}\lp q\cdot\La\rp^AG^S+\al_{3MI}\lp -\frac{1}{4}\bar{D}^2D^\al U^M+\lp q\cdot \Upsilon^\al\rp^M\rp W^I_\al\right.\right.\non\\
& \qquad\left.\left.+\al_{5XZ}\lp -\frac{1}{4}\bar{D}^2\Xi^X\rp E^Z\rp\rs\non\\
=& \int d^4xd^4\t\lp\lp\al_{2IN}q^I_M-\al_{3MI}q^I_N\rp U^MH^N-\lp\al_{4SA}q^S_X-\al_{5XZ}q^Z_A\rp\Xi^XF^A\rp\non\\
& \quad+\operatorname{Re}\ls i\int d^4xd^2\t\lp\lp\al_{1AS}q^A_I+\al_{2IM}q^M_S\rp\La^IG^S+\lp\al_{3MI}q^M_S+\al_{4SA}q^A_I\rp\Upsilon^{S\,\al}W^I_\al\rp\rs.
\end{align}
Here we have used eqn. (\ref{eq:SuperspaceMeasures}) relating the measures $d^4\t$ and $d^2\t$, the superspace analog of integrations by parts, and the Bianchi identities (\ref{E:superBI}).  We can immediately see that the conditions for gauge invariance
are precisely those found for the invariance of the bosonic action (cf. eqn. \ref{eq:QuadraticCSVariation}).

\subsubsection{Cubic Chern-Simons Terms}

Similarly we can supersymmetrize the cubic Chern-Simons action (\ref{eq:S2CS}).  First we have to make a couple of definitions.  Let
\be
\widehat{\Phi}^A=\frac{\Phi^A+\ov{\Phi}^A}{2},\qquad\widehat{E}^Z=\frac{E^Z+\ov{E}^Z}{2}=q^Z_A\widehat{\Phi}^A.
\ee
We also define an operator
\be
\Om(U,\Psi)=D^\al U\Psi_\al+\bar{D}_{\dot\al}U\ov{\Psi}^{\dot\al}+\hlf U\lp D^\al\Psi_\al+\bar{D}_{\dot\al}\ov{\Psi}^{\dot\al}\rp,
\ee
which takes as arguments a real superfield $U$ and a chiral spinor superfield $\Psi$, and returns a real superfield.  This operator has some nice properties.  In particular,
\bea
-\frac{1}{4}\bar{D}^2\Om(U,\Psi) &=& \lp -\frac{1}{4}\bar{D}^2D^\al U\rp\Psi_\al-\frac{1}{8}\bar{D}^2\ls U\lp D^\al\Psi_\al-\bar{D}_{\dot\al}\ov{\Psi}^{\dot\al}\rp\rs,\\
-\frac{1}{4}D^2\Om(U,\Psi) &=& \lp -\frac{1}{4}D^2\bar{D}_{\dot\al}U\rp\ov{\Psi}^{\dot\al}+\frac{1}{8}D^2\ls U\lp D^\al\Psi_\al-\bar{D}_{\dot\al}\ov{\Psi}^{\dot\al}\rp\rs.
\eea
Also,
\be
U_1\Om(U_2,\Psi)+U_2\Om(U_1,\Psi)=D^\al\lp U_1U_2\Psi_\al\rp+\bar{D}_{\dot\al}\lp U_1U_2\ov{\Psi}^{\dot\al}\rp,
\ee
and if we define $\Psi_i^\al=-\frac{1}{4}\bar{D}^2D^\al U_i$, then
\be
\Om(U_1,\Psi_2)-\Om(U_2,\Psi_1)=-\frac{1}{8}D^\al\bar{D}^2\lp U_1D_\al U_2-U_2D_\al U_1\rp-\frac{1}{8}\bar{D}_{\dot\al}D^2\lp U_1\bar{D}^{\dot\al}U_2-U_2\bar{D}^{\dot\al}U_1\rp,
\ee
and
\be
U_1\Om(U_2,\Psi_3)+U_2\Om(U_3,\Psi_1)+U_3\Om(U_1,\Psi_2)=D^\al\lp\cdots\rp+\bar D_{\dot\al}\lp\cdots\rp,
\ee
where we won't need the explicit form of the omitted terms $\lp\cdots\rp$ but only the fact that the right hand side is a total superspace derivative and, therefore, vanishes when integrated over $\int d^4xd^4\t$.

With these definitions, one can write the supersymmetrized Chern-Simons action as
\begin{align}
\label{eq:S2SCS}
S_{2,SCS}=& \int d^4xd^4\t\left[\al_{2ABM}\widehat{\Phi}^AF^BH^M+\al_{4IZM}V^I\widehat{E}^ZH^M+\al_{5IAJ}V^I\Om(F^A,W^J)\right.\non\\
& \qquad\left. +\al_{7MAB}F^A\Om(F^B,\Sigma^M)-\al_{8SZA}X^S\widehat{E}^ZF^A\right]+\operatorname{Re}\ls i\int d^4xd^2\t\lp\al_{1AZS}\Phi^AE^ZG^S\right.\right.\non\\
& \qquad\left.\left. +\al_{3AIJ}\Phi^AW^{I\,\al}W^J_\al+\al_{6MZI}E^Z\Sigma^{M\,\al}W^I_\al+\al_{9XZZ'}\G^XE^ZE^{Z'}\vphantom{\int}\rp\rs
\end{align}

After some manipulations, its variation has the form
\begin{align}
\d S_{2,SCS}=& \int d^4xd^4\t\left[\lp\al_{2BAM}q^B_I+\al_{4IZM}q^Z_A-\al_{5IAJ}q^J_M\rp\frac{\La^I+\ov{\La}^I}{2}F^AH^M\right.\non\\
& \quad\left. +\lp\al_{4IZN}q^I_M-\al_{6MZI}q^I_N\rp U^M\widehat{E}^ZH^N\right.\non\\
& \quad\left. +\lp\al_{5JAI}q^J_M-\al_{6MZI}q^Z_A+2\al_{7MZB}q^B_I\rp U^M\Om(F^A,W^I)\right.\non\\
& \quad\left. +\lp\al_{7MAB}q^M_S-\al_{8SZA}q^Z_B\rp F^A\Om(F^B,\Upsilon^S)-\lp\al_{8SZA}q^S_X-2\al_{9XZZ'}q^{Z'}_A\rp\Xi^X\widehat{E}^ZF^A\vphantom{\frac{\La^I+\ov{\La}^I}{2}}\right] \non\\
& +\operatorname{Re}\ls i\int d^4xd^2\t\lp\lp\al_{1AZS}q^A_I+\al_{4IZM}q^M_S\rp\La^IE^ZG^S\right.\right.\non\\
& \qquad\left.\left. +\lp\al_{3AJK}q^A_I+\al_{5IAJ}q^A_K\rp\La^IW^{J\,\al}W^K_\al+\lp\al_{6MZI}q^M_S+\al_{8SZA}q^A_I\rp E^Z\Upsilon^{S\,\al}W^I_\al\rp\vphantom{\int}\rs.
\end{align}
We see that the conditions for gauge invariance are again precisely (\ref{eq:CubicInv1}) and (\ref{eq:CubicInv2}), as in the bosonic case.

We now have all the details needed to write down the four-dimensional $N=1$ off-shell supersymmetrization of the eleven-dimensional Chern-Simons term.  It will be given by (\ref{eq:S2SCS}), with the coefficients $\al$ given by (\ref{eq:11Dalphas}).

\subsection{Descent Formalism}

We would now like to imitate the bosonic descent formalism and unify the cases above.  Thus we write the action in general as
\be
\label{eq:DescentSCS}
S_{SCS}=\int d^4xd^4\t\lp V^Ic_{3I}-X^Sc_{1S}\rp+\operatorname{Re}\ls i\int d^4xd^2\t\lp\Phi^Ac_{4A}+\Sigma^{M\,\al}c_{2M\,\al}+\G^Xc_{0X}\rp\rs.
\ee
Here $c_{3I}$ and $c_{1S}$ are real superfields, $c_{4A}$ and $c_{0X}$ are chiral superfields, and $c_{2M}$ is a chiral spinor superfield.  All of these are built out of the field strengths $E^Z$, $F^A$, $W^I$, $H^M$, and $G^S$.

Explicitly for the cases above, we have for the linear Chern-Simons action,
\be
c_{0X}=\al_X,
\ee
with the other $c$'s vanishing.  For the quadratic Chern-Simons action we have
\bea
c_{4A} &=& \al_{1AS}G^S,\non\\
c_{3I} &=& \al_{2IM}H^M,\non\\
c_{2M\,\al} &=& \al_{3MI}W^I_\al,\\
c_{1S} &=& \al_{4SA}F^A,\non\\
c_{0X} &=& \al_{5XZ}E^Z.\non
\eea
And for the cubic action,
\bea
c_{4A} &=& \al_{1AZS}E^ZG^S+\al_{3AIJ}W^{I\,\al}W^J_\al+\frac{i}{4}\al_{2ABM}\bar{D}^2\lp F^BH^M\rp,\non\\
c_{3I} &=& \al_{4IZM}\widehat{E}^ZH^M+\al_{5IAJ}\Om(F^A,W^J),\non\\
c_{2M\,\al} &=& \al_{6MZI}E^ZW^I_\al+\frac{i}{2}\al_{7MAB}\bar{D}^2\lp F^AD_\al F^B\rp,\\
c_{1S} &=& \al_{8SZA}\widehat{E}^ZF^A,\non\\
c_{0X} &=& \al_{9XZZ'}E^ZE^{Z'}.\non
\eea

For the general action (\ref{eq:DescentSCS}), invariance under variation requires
\begin{align}
\label{eq:SuperDescentRelations}
\begin{array}{lclccl}
& 0 &=& -\frac{1}{4}\bar{D}^2c_{3I} & - & q^A_Ic_{4A},
\cr
& 0 &= & \frac{D^\al c_{2M\,\al}-\bar{D}_{\dot\al}\ov{c}_{2M}^{\dot\al}}{2i} & + & q^I_Mc_{3I},
\cr
& 0 &=& -\frac{1}{4}\bar{D}^2D_\al c_{1S} & - & q^M_Sc_{2M\,\al},
\cr
& 0 &=& \frac{c_{0X}-\ov{c}_{0X}}{2i} & + & q^S_Xc_{1S}.
\end{array}
\end{align}
Again we see the appearance of the same operators.  We can also verify that for the linear, quadratic, and cubic cases above, imposing (\ref{eq:SuperDescentRelations}) is equivalent to the conditions on the $\al$'s and $q$'s that were already deduced.

\section{Prospects}

The aim of our current program is to describe the actions appearing in a supersymmetric Kaluza-Klein compactification of ten-dimensional type II theory or M-theory involving massless fields and an infinite tower of massive fields in a closed form. In recent times it has become evident that particularly the massive states include a host of physical information, such as the appearance of a new superpotential describing their interactions \cite{Becker:2014rea}.

In this paper, we have taken
a step in the direction of constructing these actions
by embedding the Abelian tensor hierarchy appearing in such reductions into four-dimensional, $N=1$ superspace
and explicitly presenting standard kinetic actions as integrals of gauge invariant chiral quantities over half of superspace or real quantities over all of superspace. We also constructed Chern-Simons-type actions which are supersymmetric in the usual way but which are only
gauge invariant after combining many terms and integrating over superspace. As we have stated, these models are inspired by but not identical to the embedding of a higher dimensional antisymmetric tensor field into $d$-dimensional superspace ($d=4$ is the example we focused on) because it has additional bosonic components needed to complete the supersymmetry multiplet.

Embedding this Abelian tensor hierarchy into superfield supergravity is non-trivial and we propose to proceed in two steps. The first step is to gauge the hierarchy with respect to the vector-like components of the dimensionally reduced metric. In a forthcoming paper \cite{BBLR2} we do this by coupling this Abelian model to non-abelian gauge fields.
The second step is to reconcile the component field mismatch alluded to above. A comparison of the components of 11D supergravity to those of the hierarchy shows that there are (at least) the $35+7$ superfluous scalars coming from the scalar and two-form multiplets, respectively as the bosonic partners required to complete the multiplet. On the other hand, the remaining supergravity components have not yet been accounted for and it is known from previous work \cite{Linch:2002wg,Gates:2003qi} that including these superspin-$\tfrac32$ and -$1$ multiplets has the potential to resolve this mismatch.
Including the coupling to these fields is work currently in progress \cite{BBLR3}.  The goal ultimately is to the embed 
the action eqn. (4.1) of ref. \cite{Becker:2014uya} in four-dimensional, $N=1$ superspace in order to learn about quantum corrections of M-theory in terms of powerful non-renormalization theorems in four dimensional superspace.

A natural toy model for eleven-dimensional supergravity is 5D, $N=1$ supergravity. It contains a ``graviphoton'' analogous to the M-theory three-form for which one can write a Chern-Simons action.
A natural thing to do, therefore, is to extend the program to include 5D, $N=1$ superspace \cite{Kuzenko:2005sz} and relate it to the supergravity theory of ref. \cite{Kuzenko:2008wr, Kuzenko:2007cj}. 
Alternatively, one can attempt to increase the amount of manifest supersymmetry to 6D, $N=(1,0)$ leaving only five additional directions and six non-linear supersymmetries. The curved superspace for such an extension was constructed in \cite{Linch:2012zh} and an action was proposed based on that of ref. \cite{Sokatchev:1988aa}. 
The action was recently reduced to 4D, $N=1$ superspace notation in ref. \cite{Abe:2015bqa,Abe:2015yya}. 
This 4D, $N=1$ description of 6D, $N=(1,0)$ supergravity and related results may prove useful in the construction of the eleven-dimensional action.

\section*{Acknowledgements}

We thank Daniel Butter, Stephen Randall, and the participants of the String-Math 2015 conference, where part of this work was presented, for discussions and interesting comments.
This work was supported by NSF grants
PHY-1214333 and PHY-1521099.


\appendix

\section{Conventions}
\label{app:Conventions}
In this appendix, we collect some oft-used identities satisfied by the four-dimensional, $N=1$ superspace covariant derivatives. Our conventions are those of \cite{Wess:1992cp} (which are closely related to those of \cite{Buchbinder:1998qv}).

The basic identities satisfied by the superspace covariant derivatives are
\be
\left\{ D_\al,\bar{D}_{\dt\al}\right\}=-2i\s^a_{\al\dt\al}\p_a,\qquad
\left\{D_\al,D_\beta\right\}=0=\left\{\bar{D}_{\dt\al},\bar{D}_{\dt\beta}\right\},
\ee
with $\s^a$ the usual Pauli matrices. The (flat) spacetime indices will be denoted by lowercase Latin letters $a,b, \dots = 0,1,2,3$. Chiral and anti-chiral spinor indices are denoted by Greek letters taking two values $\alpha,\beta, \dots =1,2$ and $\dot \alpha, \dot \beta, \dots = 1,2$.

Manipulating these fundamental $D$-algebra rules results in the following list of useful relations:
\begin{subequations}
\begin{align}
\label{E:chiralOp}
\bar D_{\dt \alpha} \bar D^2 = 0
~~~&,~~~
D_{\alpha} D^2 = 0
\\
[D^2, \bar D_{\dt \alpha}] = -4i\s^a_{\al\dt\al}\p_aD^\alpha
~~~&,~~~
[\bar D^2, D_{\alpha}] = 4i \s^a_{\al\dt\al}\p_a \bar D^{\dt \alpha}
\\
\label{E:realOp}
D^\alpha \bar D^2 D_\alpha  = \bar D_{\dt \alpha} D^2 \bar D^{\dt \alpha}
~~~&,~~~ [D^2, \bar D^2] = -4i \s^a_{\al\dt\al}\p_a [D^\alpha, \bar D^{\dt \alpha}]
\\
\Box = -\frac18 D^\alpha \bar D^2 D_\alpha  & +  \frac1{16} D^2 \bar D^2 +  \frac1{16} \bar D^2 D^2
\\
\bar D^2 D_\alpha \bar D^2 = 0
~~~&,~~~ D^2 \bar D_{\dot \alpha} D^2 = 0
.
\end{align}
\end{subequations}
These identities are crucial to our analysis and will be used repeatedly throughout the paper.

The measures on superspace are given in terms of super-covariant derivatives by
\be
\label{eq:SuperspaceMeasures}
d^2\t=-\frac{1}{4}D^2,\quad d^2\bar{\t}=-\frac{1}{4}\bar{D}^2,\quad d^4\t=\frac{1}{16}D^2\bar{D}^2.
\ee
When appearing integrated, it is implied that the result is projected onto the $\theta = 0 =\bar \theta$ subspace. For example, the chiral integral $\int d^2\theta W = -\frac{1}{4}D^2 W \Big|$ where as is standard in the superspace literature, we use the notation $(\dots)\Big|$ to indicate that $(\dots)$ is to be evaluated on the $\theta = 0 =\bar \theta$ subspace.

We
use the $Spin(3,1)\cong SL(2;\mathbb C)$ invariant $\epsilon$ and its conjugate to define
\be
\lp\ov{\s}^a\rp^{\dot\al\al}=\e^{\al\beta}\e^{\dot\al\dot\beta}\s^a_{\beta\dot\beta}.
\ee
Together with the original Pauli matrices, these satisfy $\sigma_a \bar \sigma_b + \sigma_a \bar \sigma_b = 2\eta_{ab}$ and $\bar \sigma_a\sigma_b + \bar \sigma_a \sigma_b = 2\eta_{ab}$. The opposite signs define the spin matrices which we normalize by
\be
\lp\s_{ab}\rp_\al^{\hph{\al}\beta}=\frac{1}{4}\lp\s_a\ov{\s}_b-\s_b\ov{\s}_a\rp_\al^{\hph{\al}\beta},
\ee
and
\be
\lp\ov{\s}_{ab}\rp^{\dot\al}_{\hph{\dot\al}\dot\beta}=\frac{1}{4}\lp\ov{\s}_a\s_b-\ov{\s}_b\s_a\rp^{\dot\al}_{\hph{\dot\al}\dot\beta}.
\ee
These matrices are symmetric when the upper spinor index is lowered (or vice versa).

\end{document}